\documentclass[twocolumn,pra,superscriptaddress]{revtex4}
\usepackage{color}
\usepackage{amsmath}
\usepackage{amssymb}
\usepackage{graphicx}
\usepackage{esint}
\usepackage{mathrsfs}

\begin{document}

\preprint{APS/123-QED}

\title{Quantum versus Classical Regime in Circuit Quantum Acoustodynamics}
\author{Gang-hui Zeng}
\thanks{G.H.Z. and Y.Z. contributed equally to this work}
\affiliation{Key Laboratory of Low-Dimensional Quantum Structures and Quantum Control of Ministry of Education, Key Laboratory for Matter Microstructure and Function of Hunan Province, Department of Physics and Synergetic Innovation Center for Quantum Effects and Applications, Hunan Normal University, Changsha 410081,
China}

\author{Yang Zhang}
\thanks{G.H.Z. and Y.Z. contributed equally to this work}
\affiliation{Institute of Microelectronics, Tsinghua University, Beijing 100084,
China}

\author{Aleksey N. Bolgar}
\affiliation{Moscow Institute of Physics and Technology, Institutskiy Pereulok 9, Dolgoprudny 141701, Russia}

\author{Dong He}
\affiliation{Key Laboratory of Low-Dimensional Quantum Structures and Quantum Control of Ministry of Education, Key Laboratory for Matter Microstructure and Function of Hunan Province, Department of Physics and Synergetic Innovation Center for Quantum Effects and Applications, Hunan Normal University, Changsha 410081,
China}

\author{Bin Li}
\affiliation{Institute for Quantum Information State Key Laboratory of High Performance Computing, Changsha 410081,China}

\author{Xin-hui Ruan}
\affiliation{Key Laboratory of Low-Dimensional Quantum Structures and Quantum Control of Ministry of Education, Key Laboratory for Matter Microstructure and Function of Hunan Province, Department of Physics and Synergetic Innovation Center for Quantum Effects and Applications, Hunan Normal University, Changsha 410081,
China}

\author{Lan Zhou}
\affiliation{Key Laboratory of Low-Dimensional Quantum Structures and Quantum Control of Ministry of Education, Key Laboratory for Matter Microstructure and Function of Hunan Province, Department of Physics and Synergetic Innovation Center for Quantum Effects and Applications, Hunan Normal University, Changsha 410081,
China}
\author{Le-Mang Kuang}
\affiliation{Key Laboratory of Low-Dimensional Quantum Structures and Quantum Control of Ministry of Education, Key Laboratory for Matter Microstructure and Function of Hunan Province, Department of Physics and Synergetic Innovation Center for Quantum Effects and Applications, Hunan Normal University, Changsha 410081,
China}
\author{Oleg V. Astafiev}
\email{oleg.astafiev@rhul.ac.uk}
\affiliation{Skolkovo Institute of Science and Technology, Nobel str. 3, Moscow, 143026, Russia}
\affiliation{Moscow Institute of Physics and Technology, Institutskiy Pereulok 9, Dolgoprudny 141701, Russia}
\affiliation{Royal Holloway, University of London, Egham Surrey TW20 0EX, United Kingdom}
\affiliation{National Physical Laboratory, Teddington, TW11 0LW, United Kingdom}

\author{Yu-xi Liu}
\email{yuxiliu@mail.tsinghua.edu.cn}
\affiliation{Institute of Microelectronics, Tsinghua University, Beijing 100084,
China}
\affiliation{Frontier Science Center for Quantum Information, Beijing, China}

\author{Z. H. Peng}
\email{zhihui.peng@hunnu.edu.cn}
\affiliation{Key Laboratory of Low-Dimensional Quantum Structures and Quantum Control of Ministry of Education, Key Laboratory for Matter Microstructure and Function of Hunan Province, Department of Physics and Synergetic Innovation Center for Quantum Effects and Applications, Hunan Normal University, Changsha 410081,
China}
\date{\today}

\begin{abstract}
We experimentally study a circuit quantum acoustodynamics system, which consists of a superconducting artificial atom, coupled to both a two-dimensional surface acoustic wave resonator and a one-dimensional microwave transmission line. The strong coupling between the artificial atom and the acoustic wave resonator is confirmed by the observation of the vacuum Rabi splitting at the base temperature of dilution refrigerator. We show that the propagation of microwave photons in the microwave transmission line can be controlled by a few phonons in the acoustic wave resonator. Furthermore, we demonstrate the temperature effect on the measurements of the Rabi splitting and temperature induced transitions from high excited dressed states. We find that the spectrum structure of two-peak for the Rabi splitting could become into those of several peaks under some special experimental conditions, and gradually disappears with the increase of the environmental temperature $T$. The continuous quantum-to-classical crossover is observed around the crossover temperature $T_{c}$, which is determined via the thermal fluctuation energy $k_{B}T$ and the characteristic energy level spacing of the coupled system. Experimental results agree well with the theoretical simulations via the master equation of the coupled system at different effective temperatures.
\end{abstract}

\maketitle


\section{INTRODUCTION} \label{Sec.III}
The surface acoustic waves (SAWs) are electro-mechanical vibration waves, which are usually produced and detected by interdigital transducers (IDTs) ~\cite{Datta1986,Wakana2007,Morgan2007}, and propagate along the surface of piezoelectric material. It has $10^5$ times reduction in the propagation speed compared to electromagnetic signals (typically $3000$ m/s for SAW in solid in contrast to $3.0\times10^8$ m/s for electromagnetic wave), or say, it has $10^5$ times shorter wavelengths for the same frequencies of electromagnetic signals. Thus, SAW devices play a very important role in modern electronics including signal processing for radar and mobile telephone systems~\cite{Morgan2007,Kockum2014,Guo2017}. A wide variety of acoustic devices, e.g., delay lines, bandpass filters, micro-resonators, matched filters, and sensors~\cite{Kondalkar2018,Duquesne2019,Le2019}, have been developed and are extensively used in electronic systems. In particular, SAW resonators, which are micro-electromechanical devices, are widespread in the telecommunication and sensor industries.

People have been explored new materials to fabricate SAW devices~\cite{Jin2013,Thalmeier2010} or finding new mechanism to couple the SAW devices with other systems, such that more applications of the SAW devices can be found~\cite{Thevenard2013,Kulin2020} or the SAW devices can more easily act as a platform for lab-on-a-chip systems. For example, the interaction between the acoustic wave and the fluid films~\cite{Rambach2016} could result in many potential applications in biomedicine and biotechnology. The acousto-optic interaction provides wide possibilities for light control and probe of acoustic waves~\cite{Sun2016}. In particular, bulk wave acousto-optic devices have been found numerous applications in light deflectors, modulators, and tunable filters~\cite{White1972,Shi2013,Euchner2012}.

With the rapid progress of quantum science and technology, nanomechanical resonators are recently proposed to be coupled to quantum systems. This might open up new applications of nanomechanical devices to quantum technology~\cite{Kockum2014,Guo2017,Gustafsson2014,Aref2016,Manenti2017}, including quantum signal processing, quantum memories, time delays, quantum transducers, quantum metrology, and single-phonon detection. Superconducting quantum circuit (or superconducting qubit)~\cite{Clarke2008,You2011,Devoret2013,Gu2017} is one of the most promising scalable solid-state quantum information devices~\cite{Liu2017,Rosenberg2017,Dunsworth2018}. The interaction between superconducting quantum circuits and phonons at the quantum level is called as circuit quantum acoustodynamics (cQAD)~\cite{Noguchi2017,Manenti2017}, which is similar to the circuit (cavity) quantum electrodynamics~\cite{Wallraff2004,Blais2004,Gu2017} for the interaction between superconducting quantum circuits (atoms) and microwave photons (photons).

Experimental progress for cQAD has been made in recent years. Quantum ground state and single-phonon control for local vibration quanta in a nanomechanical resonator with microwave frequency were achieved by virtue of a superconducting qubit~\cite{Connell2010}. The coupling between travelling SAW phonons and superconducting quantum circuits was recently observed~\cite{Gustafsson2014}. Furthermore, a superconducting qubit strongly coupled to bulk acoustic wave resonators~\cite{Chu2017}, single-mode SAW resonators~\cite{Bolgar2018,Satzinger2018}, multi-mode SAW resonators~\cite{Lehnert2018}, a phononic crystal~\cite{Bolgar2020} were also demonstrated, respectively. The single-phonon acoustic Stark shifts were observed by coupling a superconducting qubit to a SAW resonator in either weak~\cite{Manenti2017} or strong dispersive regime~\cite{Sletten2019}. The multi-phonon Fock states were created and controlled in bulk acoustic wave resonators by a superconducting qubit~\cite{Chu2018}. Phonon-mediated quantum state transfer and remote qubit entanglement were realized by a SAW resonator~\cite{Bienfait2019}.

Most of superconducting qubits in these cQAD experiments are coupled to either a phononic resonator or a phononic waveguide. In the experiment~\cite{Manenti2017}, a superconducting qubit is coupled to both a SAW resonator and a coplanar waveguide resonator, which is employed for
independent dispersive qubit readout. So far, all of these experiments have been performed around $10\sim 40$ mK, which are typical base temperature of a dilution refrigerator, thus the effect of thermal fluctuations on the coherence of the cQAD system is negligibly small. In other words, thermal fluctuations effect on cQAD system due to high temperatures has not been studied, especially, the quantum-to-classical transition induced by thermal fluctuations in cQAD system remains an open question.

In this work, differently from the experiment~\cite{Manenti2017}, we develop a cQAD system, which consists of a transmon qubit coupled to both a two-dimensional (2D) SAW resonator and a one-dimensional ($1$D) microwave transmission line. The strong coupling between the transmon qubit and the SAW resonator is shown by the energy splitting of the anticrossing in the transmission spectrum through the SAW resonator. We also find that the propagating microwave photons in the transmission line can be controlled by phonons inside the $2$D SAW resonator. In particular, we show the quantum-to-classical transition with the increase of the environmental temperature by observing the variations of the transmission spectra through the interaction system of the transmon qubit and the $2$D SAW resonator. The effective crossover temperature corresponding to the transition regime is estimated by thermal energy and the energy level spacing of the system.

The remainder of this paper is organized as follows: in Sec.~\ref{Sec.setup}, we present the experimental setup and measurement approach. We also experimentally characterize the SAW resonator and the transmon qubit. In Sec.~\ref{model}, we give the theoretical model to describe the interaction system between the transmon qubit and the SAW resonator. In Sec.~\ref{Sec.experiment}, the acoustic Stark shift in the weak dispersive regime is observed and used to calibrate the average phonon number inside the SAW resonator. We show that the propagation microwave photons in the transmission line can be controlled by phonons inside the SAW resonator. Then, we demonstrate that the increase of the effective environmental temperature leads to the quantum-to-classical transition of the cQAD system. In Sec.~\ref{numberical}, we theoretically analyze the experimental results by numerically solving the master equation with parameters given in our cQAD system. Results are finally summarized in Sec.~\ref{conclusion}.

\section{EXPERIMENTAL SETUP AND CHARACTERISTIC FREQUENCIES}\label{Sec.setup}

\begin{figure}[ptb]
\includegraphics[scale=0.08,clip]{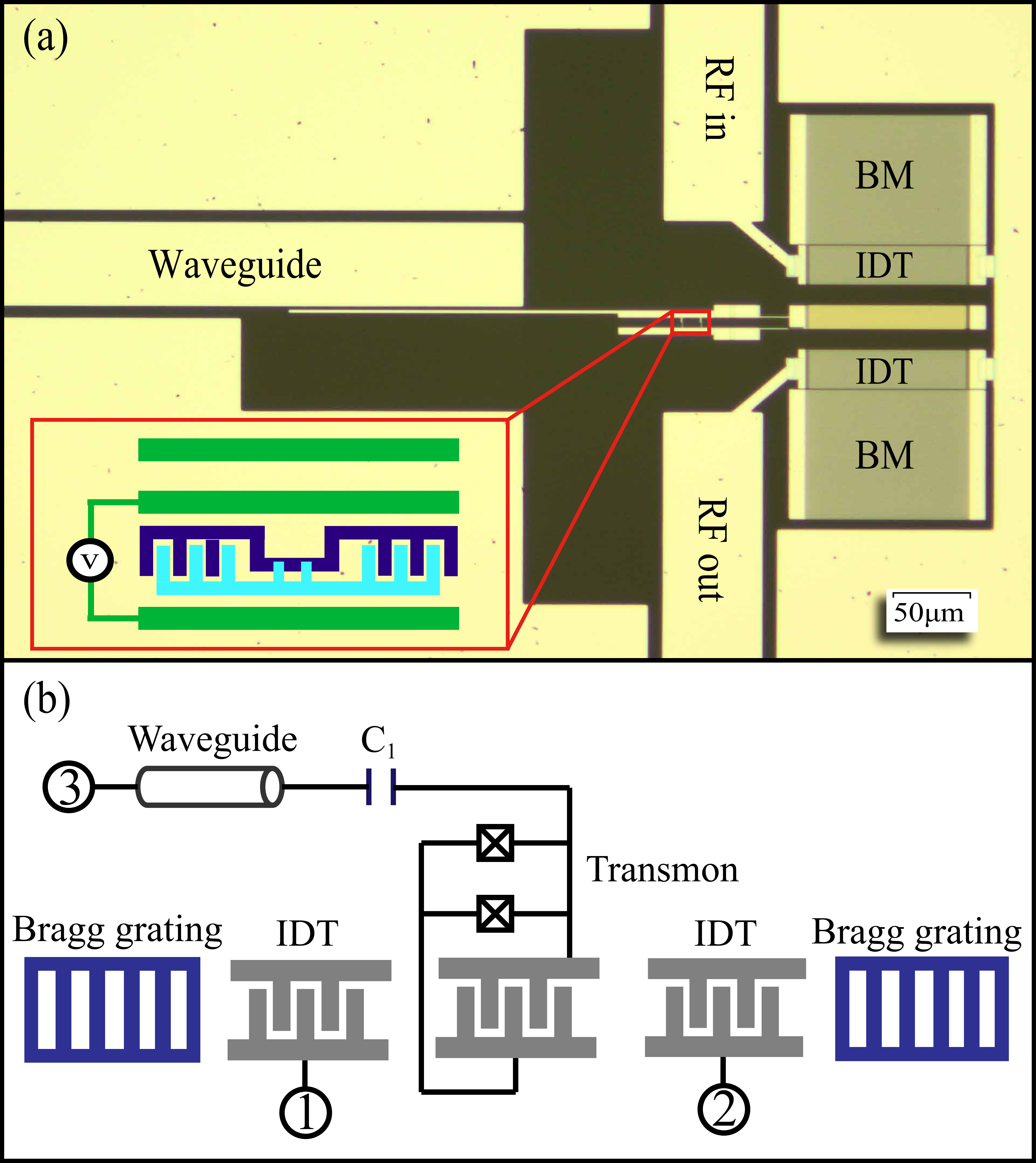}
 \caption{(Color online) Circuit quantum acoustodynamics device. (a) Optical micrograph of the device. A transmon (inset) is strongly coupled to the 2D SAW resonator via the interdigital transducer (IDT), and simultaneously coupled to a coplanar waveguide transmission line via a capacitor. BM is an abbreviation of the Brag Mirror.  (b) An equivalent electrical circuit of the device. A SAW resonator is excited and probed via two interdigital transducers (IDTs) connected to ports $1$ and $2$.  The coplanar waveguide transmission line is used to measure the transmon via port $3$. The Brag gratings act as the Brag mirrors.}

\label{Figure1}
\end{figure}

Our device, as shown in Fig.~\ref{Figure1}(a), is fabricated on a quartz substrate. The equivalent electrical circuit in Fig.~\ref{Figure1}(a) is schematically shown in  Fig.~\ref{Figure1}(b).  The cQAD device is formed by a tunable SQUID-based transmon qubit strongly coupled to a 2D SAW resonator via a shunted IDT~\cite{Bolgar2018}. The electrodes of IDT are design to position at the antinodes of the coupled acoustic wave in the resonator. Simultaneously, the transmon qubit is also capacitively coupled to $1$D coplanar waveguide transmission line, which is used to measure the qubit and the cQAD system via the port $3$. A SAW is excited and detected by two identical IDTs connected to ports $1$ and $2$. Two Brag gratings, acting as two Brag mirrors, are used to form a SAW resonator. All IDTs are formed by a periodic array of alternating stripe electrodes and have the same period $p$. The SAW travelling speed in quartz at low temperatures is about $3.16$ km/s. In our device, the periodicity of the IDT stripes is $p=980$ nm and that of the Bragg mirrors is $p/2$, which defines the SAW wavelength $\lambda=p$ and the frequency about $3.2$ GHz.  Our device is mounted at the mixing chamber of a dilution refrigerator with the base temperature $T\approx16.5$ mK. A superconducting Al shield and two $\mu-$metal shields are used to reduce the ambient magnetic field.

\begin{figure}[ptb]
	\includegraphics[scale=0.25,clip]{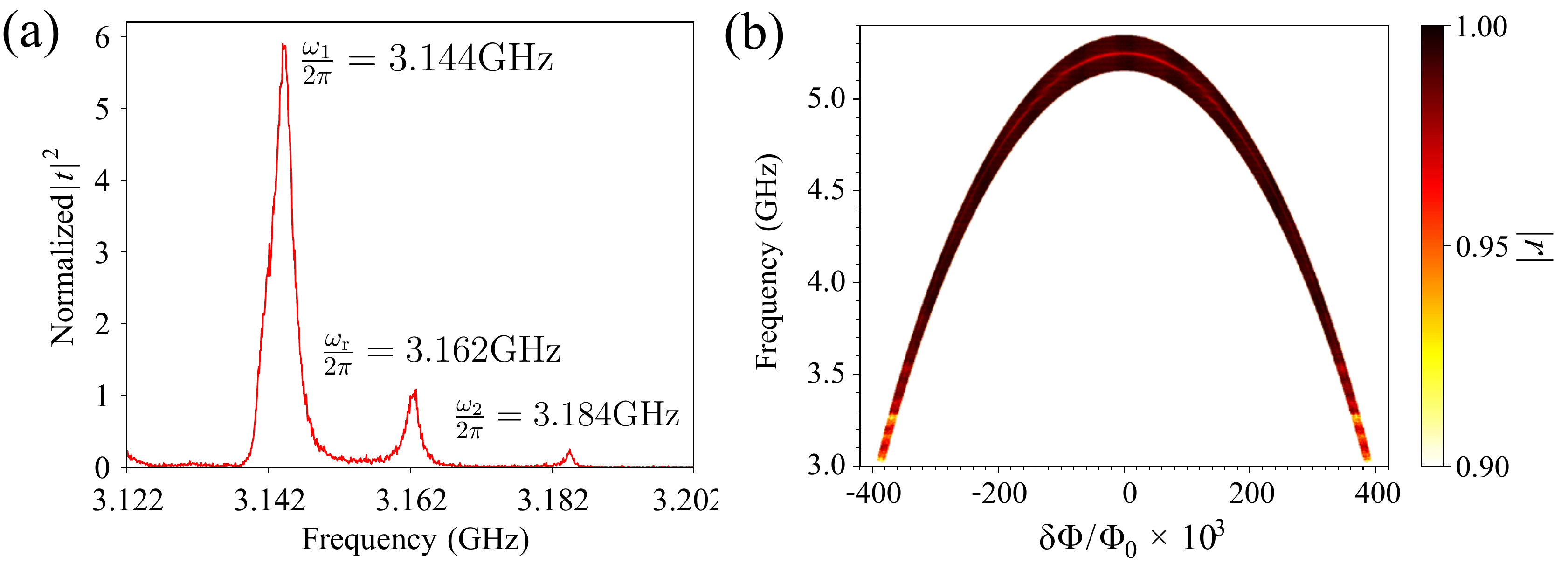}
		\caption{(Color online) (a) Normalized transmission through the SAW resonator. The transmon is coupled to the mode $\omega_r=2\pi\times3.162\,$GHz and decoupled to another two modes $\omega_{1}=2\pi\times3.144\,$GHz and $\omega_{2}=2\pi\times3.184\,$GHz. (b) Reflection spectrum of the transmon qubit vs biased flux plotted as a reflection coefficient $|r|$.}
	\label{Figure2}
\end{figure}

Our device is characterized as it is discussed below. We measure the transmission coefficient $t$ through the SAW resonator from the port $1$ to the port $2$ at the base temperature, and then obtain frequencies of the SAW resonator. Figure~\ref{Figure2}(a) shows the normalized transmission coefficient $|t|^2$ with the probing power lower than $-134$ dBm, which corresponds to the average phonon number around $1$ inside the SAW resonator. There are three resonances at the frequencies $\omega_{1}=2\pi\times3.144$ GHz, $\omega_r=2\pi\times3.162$ GHz, and $\omega_2=2\pi\times3.184$ GHz, respectively. In the experiment, we use acoustic Stark shifts, in contrast to the ac-Stark shift in the system of the circuit QED~\cite{Schuster2005,Ong2011},  to calibrate the phonon number inside the resonator, which will be further discussed later in the paper. The decay rate of the resonator mode at the frequency $\omega_r=2\pi\times3.162$ GHz is $\kappa=2\pi\times1.56$ MHz by extracting the linewidth from $|t|^2$.

The frequency of the qubit can be characterized by measuring the reflection coefficient $r$ of the microwave signal through the waveguide via the port $3$ using a vector network analyzer (VNA) at the base temperature. Figure~\ref{Figure2}(b) shows a $2$D plot of the normalized reflection coefficient $|r|$ in the frequency range $3.0\sim5.4$ GHz with the magnetic flux bias $\delta\Phi$ from $-400 \,\text m\Phi_0$ to $400\,\text m\Phi_0$, where $\Phi_0$ is the flux quantum. The transition from the ground state to the first excited state of the qubit is revealed as dips. The transition frequency $\omega_a$ at $\delta\Phi=0$ reaches the maximum value $\omega_a=2\pi\times5.260$ GHz and the extracted relaxation rate at this point is $\Gamma=2\pi\times10.48$ MHz, which is similar to that in Ref.~\cite{Bolgar2018}. By fitting the spectroscopy of the qubit with the formula $\omega_a=\sqrt{8E_{\rm C}E_{\rm J}(\Phi)}-E_{\rm C}$, we obtain the qubit charging energy $E_{\rm C}\approx2\pi\times160$ MHz, and the Josephson energy $E_{\rm J}(\Phi)=E_{\rm J,max}|\cos(\pi\Phi)/\Phi_0|$, in which $E_{\rm J,max}\approx2\pi\times22.88$ GHz. In experiment, we can adjust the transition frequency of the qubit by changing the bias flux $\Phi$, injected to the SQUID loop of the qubit.

\section{THEORETICAL MODEL}\label{model}

Our cQAD system is formed by a transmon qubit interacting with a single-mode SAW resonator, thus this system can be modeled as a two-level system interacting with a single-mode cavity, given in the cavity quantum electrodynamics system.  Below, the transmon qubit is called as the qubit for convenience. The single-mode Hamiltonian of the SAW resonator~\cite{Noguchi2017,Manenti2016} can be written as $H/\hbar=\omega_r({\hat{a}^\dagger}\hat{a}+\frac{1}{2})$, where $\hat{a}^\dagger(\hat{a})$ is the bosonic creation (annihilation) operator of the single SAW resonator mode and $\hbar$ is the reduced Plank constant. In our experiment, the mode of the SAW resonator with the frequency $\omega_r=2\pi\times3.162$ GHz  is coupled to the qubit. The Hamiltonian of the qubit can be written as $H/\hbar=\omega_a\sigma_z/2$ with the transition frequency $\omega_a$ of the qubit, $\sigma_z=|e\rangle\langle e|-|g\rangle\langle g|$ is the Pauli $Z$ operator with the ground state $|g\rangle$ and the excited state $|e\rangle$ of the qubit. The interaction Hamiltonian between the SAW resonator and the qubit is $H_{\rm int}/\hbar=g\sigma_x({\hat{a}^\dagger}+\hat{a})$ with the coupling strength $g$. Under the rotating-wave approximation, the total Hamiltonian is given as below
\begin{equation}
H/\hbar=\omega_r{\hat{a}^\dagger}\hat{a}+\frac{1}{2}\omega_a\sigma_z+g(\sigma_{+}\hat{a}+\sigma_{-}\hat{a}^\dagger),\label{Eq:1}
\end{equation}
where $\sigma_{+}=|e\rangle\langle g| \,(\sigma_{-}=|g\rangle\langle e|)$ is raising (lowering) operator of the transmon qubit. Hereafter, the state $|e,n\rangle$ ($|g,n\rangle$) denote that the qubit is in the excited (ground) state and there are $n$ phonons in the SAW resonator. The ground state $|g,0\rangle$ of the system means that the qubit is in the ground state and the SAW resonator is in the vacuum state, this state can be prepared in our experiment by cooling the system down to the base temperature of the dilution refrigerator.

\begin{figure}[ptb]
	\includegraphics[scale=0.12,clip]{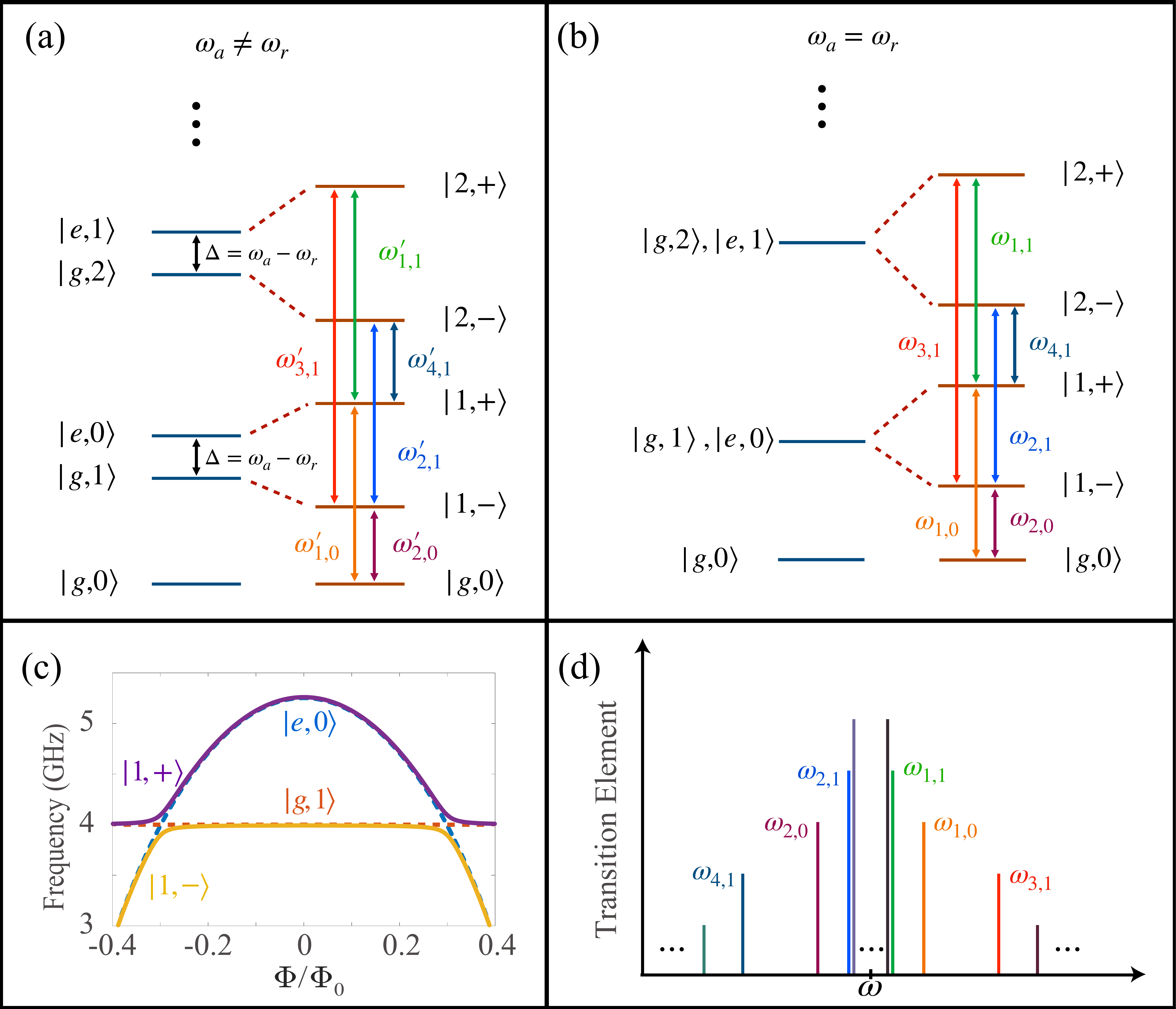}\
	\caption{(Color online) Energy spectrum of the uncoupled (left) and coupled  qubit-resonator system (right) under the nonresonant condition (a) and resonant condition (b). The degeneracy of the states $|n,e\rangle$ and $|n+1,g\rangle$ in (b) is lifted $2g\sqrt{n+1}$ by the coupling. (c) Energy diagram of the qubit (dashed blue curve $|e,0\rangle$), the SAW resonator (dashed orange curve $|g,1\rangle$), the coupled system (solid purple and yellow curve $|1,\pm\rangle$), parameters we choose here are $E_{\rm C}=2\pi \times 160$MHz, $E_{\rm J,max}=2\pi \times 22.88$GHz, $g=2\pi \times 100$MHz and $\omega_{r}=2\pi \times 4$GHz. (d) Schematic diagram for transitions corresponding to different frequencies as shown in (b).}
	\label{Figure3}
\end{figure}

The coupling between the qubit and the SAW resonator mixes the states $|g,n+1\rangle$ and $|e,n\rangle$ as two dressed states
\begin{eqnarray}
   	|n,+\rangle&=\cos\theta_{n}|g,n\rangle + \sin\theta_{n}|e,n-1\rangle, \\
	|n,-\rangle&=\sin\theta_{n}|g,n\rangle - \cos\theta_{n}|e,n-1\rangle,
\end{eqnarray}
as schematically shown in Fig.~\ref{Figure3}(a) for general case, and the corresponding eigenvalues write
\begin{equation}
E_{n,\pm}=\omega_r\left(n-\frac{1}{2}\right)\pm \frac{1}{2}\sqrt{\Delta^2+4g^2n},
\end{equation}
with $n\geq 1$, $\Delta=\omega_a-\omega_r$ and $\theta_{n}$ determined by
\begin{equation}
	\tan \left(2\theta_{n}\right)=\frac{2g\sqrt{n}}{\Delta}.
\end{equation}

When the qubit and the SAW resonator are not resonant, and in the larger detuning regime, i.e., $|\Delta=\omega_a-\omega_r|\gg g$, the Hamiltonian in Eq.~(\ref{Eq:1}) can be equivalently written
\begin{equation}
H_{\rm eff}/\hbar=\left(\omega_r+\frac{g^2}{\Delta}\sigma_z\right)\hat{a}^\dagger \hat{a}+\frac{1}{2}\left(\omega_a+\frac{g^2}{\Delta}\right)\sigma_z.
\end{equation}
In this nonresonant case, the dressed qubit level separation is given by $\widetilde{\omega_a}=\omega_a+2ng^2/\Delta+g^2/\Delta$, which depends on the number of phonons $n=\langle \hat{a}^\dagger \hat{a} \rangle$ inside the SAW resonator. We call the terms $2ng^2/\Delta$ and $g^2/\Delta$ as the acoustic Stark and Lamb shifts of the qubit frequency, in contrast to the ac Stark and Lamb shifts in the system of the cavity quantum electrodynamics, and  use Stark shift to calibrate the phonon number inside the SAW resonator induced by driving field in the experiment.

When the transmon qubit resonantly interacts with the SAW resonator, i.e., $\omega_{a}=\omega_{r}=\omega$ and $\Delta=0$ , the energy spacing becomes $\Delta_{n}=2g\sqrt{n}$, corresponding to two dressed states
\begin{equation}
|n,\pm\rangle=\frac{1}{\sqrt{2}}(|g,n\rangle\pm|e,n-1\rangle),
\end{equation}
as schematically shown in Fig.~\ref{Figure3}(b).

If a probe field is applied to the SAW resonator, in which the interaction Hamiltonian between the resonator and the driving field is proportional to $a^{\dagger}+a$, then transitions from states $|1,\pm\rangle$ to the state $|g, 0\rangle$ as well as from the states $|n+1,\pm\rangle$ to the states $|n,\pm\rangle$ can be induced. The transitions between the state $|g,0\rangle$ and the states $|1,\pm\rangle$ with frequencies $\omega+g$ and $\omega-g$ lead to the standard vacuum Rabi splitting, which can be observed with two peaks, separated by $2g$, in the absorption spectrum. This vacuum Rabi splitting corresponds to an anticrossing in the energy diagram of the coupled qubit-resonator system as shown in Fig.~\ref{Figure3}(c).

For the upper levels, there are four transitions from the states $|n+1,\pm\rangle$ to the states $|n,\pm \rangle$ for $n\geq 1$ corresponding to four frequencies
\begin{eqnarray}
\omega_{1,n}&=&\omega+g\left(\sqrt{n+1}-\sqrt{n}\right),\label{Eq:7}\\
\omega_{2,n}&=&\omega-g\left(\sqrt{n+1}-\sqrt{n}\right),\label{Eq:8}\\
\omega_{3,n}&=&\omega+g\left(\sqrt{n+1}+\sqrt{n}\right),\label{Eq:9}\\
\omega_{4,n}&=&\omega-g\left(\sqrt{n+1}+\sqrt{n}\right),\label{Eq:10}
\end{eqnarray}
here, the subscript $``i, n"$ means the $i$th transition from $|n+1,\pm\rangle$ to $|n,\pm \rangle$. $\omega_{1,n}$ and $\omega_{3,n}$ are frequencies from the state $|n+1,+\rangle$ to the states $|n,+\rangle$ and $|n,-\rangle$, respectively. However  $\omega_{2,n}$ and $\omega_{4,n}$ are frequencies from the state $|n+1,-\rangle$ to the states $|n,-\rangle$ and $|n,+\rangle$, respectively.
Thus, as schematically shown in Fig.~\ref{Figure3}(d), in the quantum case, many peaks appear at the positions $\omega\pm g$, $\omega\pm g(\sqrt{n+1}-\sqrt{n})$, and $\omega\pm g(\sqrt{n+1}+\sqrt{n})$ for all $n$ with $n\geq 1$ in the spectrum. If $n$ is a very large number such that
\begin{equation}
\omega_{2,n}-\omega_{1,n}=2g(\sqrt{n+1}-\sqrt{n})\approx \frac{g}{\sqrt{n}}\approx 0,
\end{equation}
then $\omega_{1,n}\approx \omega_{2,n}=\omega$, that is, the two peaks for frequencies $\omega_{1,n}$ and $\omega_{2,n}$ in the spectrum merge into one in the limit of the large $n$. It is not difficult to find that the separation between two successive peaks, e.g., corresponding to frequencies $\omega_{i, n+1}$ and $\omega_{i,n}$ with $i=1,\,2,\,3,\,4$, is of the order of $g/\sqrt{n}$. Thus, many peaks are not resolved in the limit of the large phonon number $n$ inside the resonator. Moreover, we can also find that the transition elements from the state $|n+1,+ \rangle$ to the state $|n, +\rangle$ or from the state $|n+1,- \rangle$ to the state $|n, -\rangle$ are proportional to $(\sqrt{n+1}+\sqrt{n})$, but the transition elements from the state $|n+1,- \rangle$ to the state $|n, +\rangle$ or from the state $|n+1,+\rangle$ to the state $|n, -\rangle$ are proportional to $(\sqrt{n+1}-\sqrt{n})$. Roughly speaking, the heights of the peaks corresponding to the frequencies $\omega_{1,n}$ and $\omega_{2,n}$ are higher than those of peaks corresponding to the frequencies $\omega_{3,n}$ and $\omega_{4,n}$ in the spectrum. Thus, in the limit of the large $n$, peaks corresponding to the frequencies $\omega_{3,n}$ and $\omega_{4,n}$ in the spectrum are not observable, and peaks corresponding to the frequencies $\omega_{1,n}$ and $\omega_{2,n}$ in the spectrum emerge into one, the system is changed from the quantum to the classical regime.

In practice, the environmental temperature of the system can be used to induce quantum-to-classical transition. For example, in our experiments, the mean thermal phonon numbers $\langle n_{\rm th}\rangle=[\exp(\hbar\omega_{r}/k_{B}T)-1]^{-1}$ for the resonator and $\langle n_{\rm th}^{\prime}\rangle=[\exp(\hbar\omega_{a}/k_{B}T)-1]^{-1}$ for the qubit are  about $\langle n_{\rm th}\rangle\approx \langle n_{\rm th}^{\prime}\rangle \approx 0$ with $\omega_r=2\pi\times3.162$ GHz  and $\omega_a=2\pi\times3\sim2\pi\times5$ GHz at the base temperature $T=16.5$ mK, and thus the cQAD system can be prepared to the ground state $|g,0 \rangle$,  the transitions mostly occur between the state $|g, 0 \rangle$ and the states $|1,\pm\rangle$ at the base temperature. In this case, transition spectra of dressed states $|1,\pm\rangle$ to the state $|g, 0 \rangle$, with two peaks separated by $2g$, can be observed. However, with the increase of the temperature, the mean thermal phonon number $\langle n_{\rm th}\rangle$ and $\langle n_{\rm th}^{\prime}\rangle$  are also increased, then more high energy levels are excited, and the spectra with multiple peak structure should be observed at the intermediate temperatures. If the temperature is further increased such that the number of the thermal phonons is very large, multiple peaks merge into two peaks, and finally one peak. This is because: (i) the lower energy levels are all saturated at the high temperature and transition can only occur from large quantum number, whose transition frequencies are densely close to the resonator frequency, as discussed above, (ii) the spectrum width of each peak becomes broader and broader with the increase of the temperature,  thus many peaks overlapped into one. The temperature effect on our experimental results will be carefully analyzed and discussed in following sections.

\section{EXPERIMENTAL RESULTS}\label{Sec.experiment}

At the base temperature $T\approx 16.5$ mK of the dilution refrigerator, the mean thermal excitation phonon numbers for both the acoustic mode frequency $\omega_{r}$  and the qubit frequency are about zero, as discussed above. Thus, the system works in the quantum regime because the effect of thermal fluctuations on the coupled system between the SAW resonator and the qubit is negligibly small and the quantum fluctuations are dominant. In our experiments, the qubit is designed to be coupled to the SAW mode with the frequency $\omega_r=2\pi\times3.162$ GHz by matching $p$ with the wavelength of the frequency $\omega_{r}$,  and decoupled from other two modes with frequencies $\omega_1=2\pi\times 3.144$ GHz and $\omega_2=2\pi\times3.184$ GHz. Note that the coupling between the SAW resonator and the qubit is usually characterized in two ways. (i) When the qubit and the SAW resonator are in the regime of large detuning, the coupling is measured by the acoustic Stark shift of the qubit frequency or the dispersive shift of the SAW resonator frequency induced by the qubit. (ii) When the qubit resonantly interacts with the SAW resonator, the coupling strength can be found by measurement of the anticrossing.

\subsection{Non-resonant case}
\begin{figure}[ptb]
	\includegraphics[scale=0.25,clip]{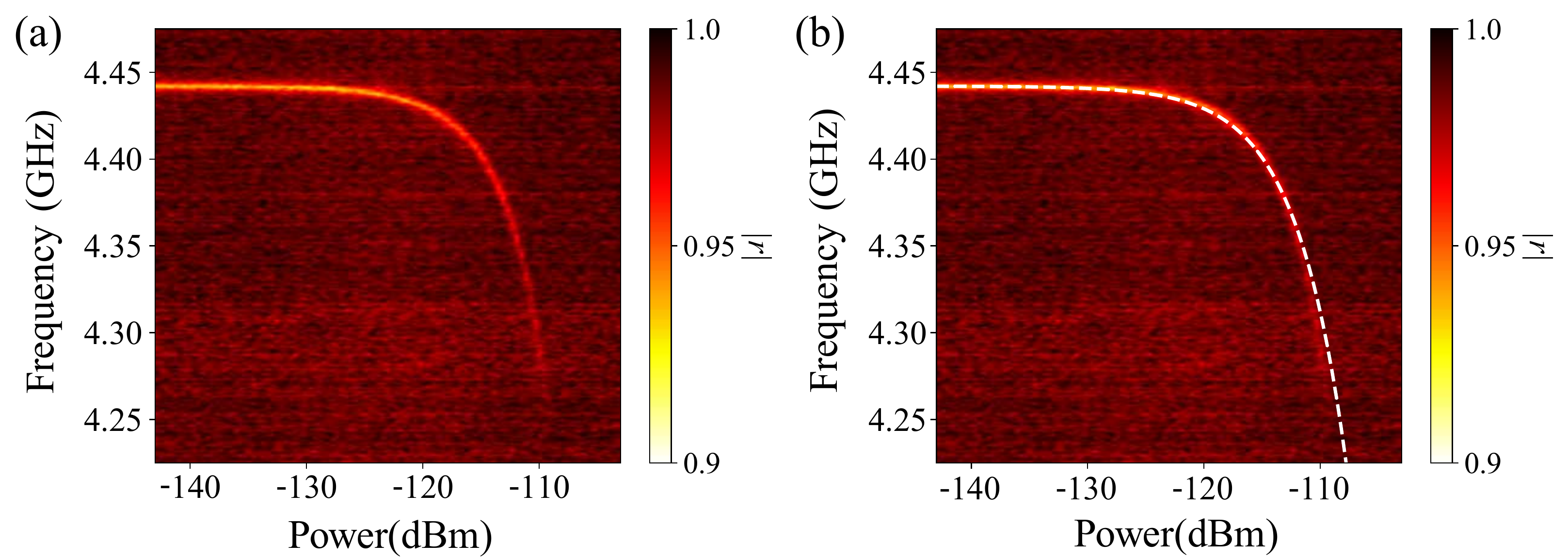}
	\caption{(Color online) (a) Spectrum of the transmon vs driving power of the SAW resonator as a reflection coefficient $|r|$. $-134\,$dBm corresponds to average phonon number $\langle a^{\dagger}a\rangle=\langle n\rangle=1$ inside the SAW resonator. (b) The spectrum with theoretically calculated transition frequency (dashed line) from ac-Stark shift.}
	\label{Figure4}
\end{figure}

We first study the large detuning case. In Refs.~\cite{Manenti2017} and~\cite{Sletten2019}, the single-phonon acoustic Stark shift in the weak and the strong dispersive regimes are observed, respectively. However, the phonon number dependence of the acoustic Stark shift from a few to large number of phonons is still missing, because it was not easy to clearly detect large qubit frequency shift with limited measurement bandwidth in the SAW resonator in previous cQAD systems. In our experiment, the qubit is coupled to the transmission line and the limitation of the measurement bandwidth is overcome. Thus, it is not difficult to detect the phonon number dependence of the acoustic Stark shift in the large phonon number range. We apply the magnetic flux bias such that the qubit works at the frequency $\omega_a=2\pi\times4.442$ GHz. In this case, the detuning between the qubit and the SAW resonator is $\Delta=2\pi\times1.282$ GHz, which is much larger than $g$. Then we measure the phonon-dependent acoustic Stark shift via the reflection spectrum of the qubit through the port $3$ using VNA. The acoustic phonon number is changed by adjusting the driving power on the SAW resonator. As shown in Fig.~\ref{Figure4}(a), we measure the phonon dependent acoustic Stark shift in the range from $-143$ dBm to $-103$ dBm, which corresponds to the average phonon number from $\langle n\rangle\sim0.1$ to $\langle n\rangle\sim1000$ inside the SAW resonator. Here, we define the phonon number operator $n=a^{\dagger} a$.  The acoustic Stark shift induced by one phonon is extracted to be $2g^2/\Delta\approx510$ KHz via the experimental data in Fig.~\ref{Figure4}(b), which means the device works in the weak dispersive regime. With the power calibration, we find that the microwave to phonon conversion efficiency of each IDT is $-8\pm3$ dB. To keep low phonon number excitation, the probing power used in following experiment is set lower than $-134$ dBm, which corresponds to the average phonon number $\langle n\rangle\sim1$ inside the SAW resonator. We note that although the qubit is weakly coupled to the $1$D transmission line, we can clearly resolve the reflection spectrum of the propagation microwave field for the average phonon number inside the SAW resonator down to $\langle n\rangle\sim 21$ because of $21\times2g^2/\Delta\ >\Gamma,\kappa$. This means that the reflection of the microwave field could be controlled by the phonons inside the SAW resonator, which is similar with the demonstration in Ref.\cite{Peng2018} that reflection of the microwave field controlled by the microwave photons inside a microwave transmission line resonator. Here, our result is the first step towards the microwave photon control by phonons in a quantum hybrid system on-chip.

 \begin{figure*}[ptb]
	\includegraphics[scale=0.13,clip]{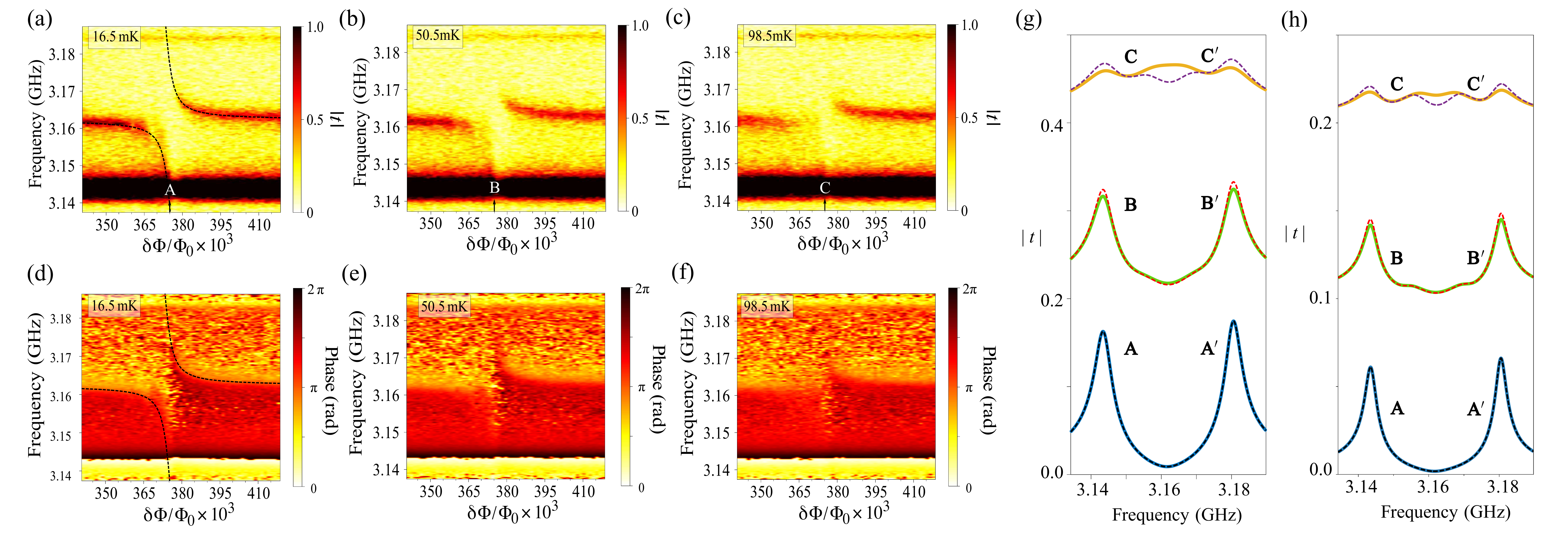}\
	\caption{(Color online) (a), (b), and (c) Normalized transmission amplitude around the SAW mode($\omega_r=2\pi\times3.162\,$GHz) as a function of $\delta\phi$ under different temperatures. The temperature of the system are $16.5\,$mK, $50.5\,$mk, $98.5\,$mk, respectively, from (a) to (c). (d), (e), and (f) Phases of the transmission spectra under different temperatures from the experimental data, the temperatures of the system are $16.5\,$mK, $50.5\,$mK, $98.5\,$mK, respectively, from (d) to (f). (g) The calculated spectra around $\omega_a=\omega_r$ with the model of the lowest five energy levels of transmon (the model of the lowest two levels of transmon), plotted with solid curves (dashed curves) for A(A$'$), B(B$'$), and C(C$'$) with experimental parameters $\kappa=2\pi\times1.56\,$MHz and $\gamma=2\pi\times7.90\,$MHz, and corresponding temperatures are  $16.5\,$mK, $50.5\,$mK, $98.5\,$mK, respectively. (h) The calculated spectra around $\omega_a=\omega_r$ with the model of the lowest five energy levels of transmon(the model of the lowest two levels of transmon) plotted with solid (dashed) curves A(A$'$), B(B$'$), and C(C$'$) with theoretically chosen parameters $\kappa=2\pi\times0.4\,$MHz and $\gamma=2\pi\times6\,$MHz, and the corresponding temperatures are $16.5\,$mK, $50.5\,$mK, $98.5\,$mK, respectively.}
	\label{Figure5}
\end{figure*}

\subsection{Resonant case}

In the resonant case, the interaction between the qubit and the SAW resonator can be studied via the transmission coefficient $t$ through the SAW resonator around the frequency $\omega_r=2\pi\times 3.162$ GHz using the VNA by varying magnetic flux bias $\delta\Phi$. As shown in Fig.~\ref{Figure5}(a) for the amplitudes of normalized transmission spectra measured at the base temperature $T\approx 16.5$ mK, in which the thermal phonon numbers for the qubit and resonator frequecies are $\langle n_{\rm th}\rangle \approx \langle n_{th}^{\prime}\rangle \approx0$, two hybridized modes of the qubit and the resonator are repelled by each other, and the anticrossing due to the vacuum Rabi splitting is clearly observed when $\omega_a=\omega_r$.   The coupling strength $g=2\pi\times18.5$ MHz is extracted by fitting the anticrossing, which is closed to the value obtained in Ref.~\cite{Bolgar2018}. $2g$ is just a frequency difference between two transition frequencies from the dressed states $|1,\pm\rangle$ to the ground state $|0,g\rangle$, and is the distance between two peaks in the transmission spectrum as shown in the curve A (or A$^{\prime}$ calculated using the model of the lowest two energy levels) calculated using the model of the lowest five energy levels of Fig.~\ref{Figure5}(g) for $\omega_{a}=\omega_{r}$, which is simulated using experimental parameters and the master equation  (see, Subsec. B of Sec.~\ref{numberical}).  The corresponding phase spectra from experimental data are presented in Fig.~\ref{Figure5}(d), in which the vacuum Rabi splitting is also observed. This result indicates that the strong coupling between the qubit and the SAW resonator is realized, i.e., $g>\Gamma, \,\kappa$.

The vacuum Rabi splitting can also be considered as the normal mode splitting for two coupled systems using the semiclassical theory. To further explore the quantum nature for the coupled system of the qubit and the SAW resonator, the phonon number dependent coupling should be observed, thus we slightly increase the temperature to, e.g., $T=50.5$  mK and $T=98.5$ mK, which correspond to the thermal phonon number $\langle n_{\rm th}\rangle=\langle n_{\rm th}^{\prime}\rangle\approx0.04 $ and  $\langle n_{\rm th}\rangle=\langle n_{\rm th}^{\prime}\rangle\approx 0.17 $ around the resonant frequency, respectively. In this case, the thermal phonon number is very small and the system still works in the quantum regime, but the temperature induced transitions from high energy levels are also involved, e.g., four lower transitions from the states $|2,\pm \rangle$ to the states $|1,\pm \rangle$. In Figs.~\ref{Figure5}(b), (c), (e) and (f), the amplitudes and the corresponding phases of the transmission spectra are measured at two temperatures.

Using experimental parameters in Figs.~\ref{Figure5}(b) and (c), the spectra corresponding to $\omega_{a}=\omega_{r}$ are simulated by the master equation (see, Subsec. B of Sec.~\ref{numberical}) for the  transmon model with the lowest five energy levels because of the weak anharmonicity and shown in solid curves B and C of Fig.~\ref{Figure5}(g) for these two temperatures. As comparison, the spectra corresponding to the transmon model with the lowest two energy levels are plotted as dashed curves B$'$ and C$'$ in Fig.~\ref{Figure5}(g). In curve C$'$, we find two more peaks appear in the middle besides the two peaks corresponding to the vacuum Rabi splitting. These two peaks correspond to the transitions from the state $|2,+\rangle$ to the state $|1,+\rangle$  and from the state $|2,-\rangle$ to the state  $|1,-\rangle$. These additional peaks show a quantum nature of the coupled system of the qubit and the SAW resonator. However, these additional peaks are difficult to distinguish in curve C because the quantum nature of the coupled system is covered by the weak anharmonicity of the transmon at high temperature. In order to further show the dissipation effect on the spectra, we present the calculated spectra in Fig.~\ref{Figure5}(h) around $\omega_{a}=\omega_{r}$ by theoretically choosing smaller dissipation parameters of the system, e.g., $\kappa=2\pi\times0.4\,$MHz and $\gamma=2\pi\times6\,$MHz. As shown in curve $C^{\prime}$ in Fig.~\ref{Figure5}(h), the multi-peak structure becomes more clear in the case of the small dissipation and large anharmonicity (two energy level model).  In our experiments, the conversion efficiency from microwave photon to phonon through the IDT is low and the transmon is weakly coupled to the transmission line. It is difficult to monitor the additional peaks either through SAW resonator or transmission line in experiments even that the dissipation rates become small. We expect the these additional peaks can be experimentally observed in an improved architecture with small dissipations,  a strong anharmonicity (e.g., a flux qubit\cite{Peng2018}), high conversion efficiency for IDT.

\section{Temperature induced quantum-to-classical transitions: Experimental and numerical results}\label{numberical}

In this section, we will systematically study the temperature effect on the cQAD system when the qubit resonantly interacts with the SAW resonator. In particular, we will study the quantum-to-classical transition induced by the temperature. To show a clear comparison for the effect of different temperatures on the quantum nature of the system, we regulate temperatures of the mixing chamber of the refrigerator such that different thermal phonon occupations for the SAW resonator and the qubit can be induced.

\subsection{Experimental results and qualitative analysis}

\begin{figure*}[ptb]
	\includegraphics[scale=0.17,clip]{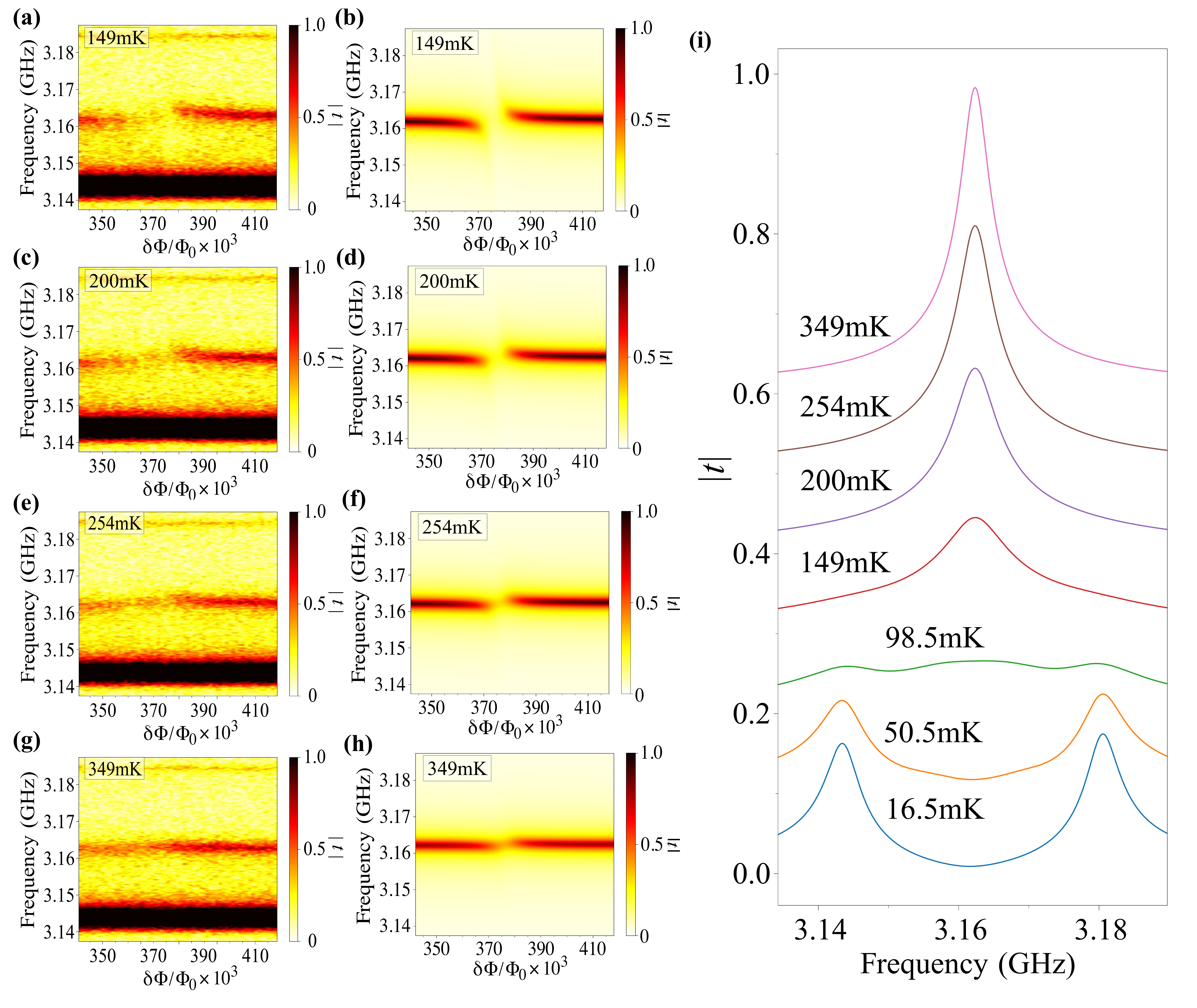}\
	
	\caption{(Color online) Normalized transmission amplitude around the SAW mode($\omega_r=2\pi\times3.162\,$GHz) as a function of $\delta\phi$ under different temperatures. (a),(c),(e) and (g) are the experimental results at $149$ mK, $200$ mk, $254$ mk, $349$ mk, respectively. The corresponding numerical simulations are shown in (b),(d),(f) and (h), respectively. (i) The calculated spectra when the SAW resonator resonantly interacts with the qubit, i.e. $\omega_{r}=\omega_{a}$, for the temperatures $16.5\,$mK, $50.5\,$mK, $98.5\,$mK, $149\,$mK, $200\,$mK, $254\,$mK and $349\,$mK, respectively.}
	\label{Figure6}
\end{figure*}

 In our experiments, we measure the transmission spectrum through the SAW resonator around the frequency $\omega_r$ by varying the temperature from $16.5$ mK to $349$ mK. Figures~\ref{Figure5}(a)-(c) show the experimental results for the temperatures $16.5$ mK, $50.5$ mK and $98.5$ mK.   Figures~\ref{Figure6}(a), (c), (e) and (g) show the experimental results when the coupled system is in different temperatures, i.e., $149$ mK, $200$ mK, $254$ mK, and $349$ mK. Figures~\ref{Figure6}(b), (d), (f) and (h) are simulation results using experimental parameters. Figure~\ref{Figure6}(i) is transmission spectra corresponding to Figs.~\ref{Figure6}(b), (d), (f) and (h) when
the qubit resonantly interacts with the resonator, i.e., $\omega_{a}=\omega_{r}$. The resonant transmission spectra corresponding to Figs.~\ref{Figure5}(a)-(c) are also plotted in Fig.~\ref{Figure6}(i) for comparison. As shown in Fig.~\ref{Figure5} and Fig.~\ref{Figure6}, with the increase of the temperature $T$, the thermal fluctuations start to play the role, two hybridized modes bend toward each other, the level repelling to the level attraction becomes obvious. When the temperature is close to $T_{c}\sim\hbar\omega_{a}/k_{B}\approx 149$ mK, with the Boltzmann constant $k_{B}$, as shown in Fig.~\ref{Figure6}(a), the two hybridized modes almost coalesce at the point of the resonant frequency $\omega_a=\omega_r=2\pi\times3.162$ GHz, the Rabi splitting almost disappears. When $T\gg T_{c}$, the thermal fluctuation of the system is dominant. The lower energy levels of the system are saturated and the energy level splitting completely disappears, and the transmission spectrum, as shown in Fig.~\ref{Figure6}(g) and the curve corresponding to the temperature $349$ mK in Fig.~\ref{Figure6}(i),  resembles that of the harmonic oscillator, and has only one peak located at the frequency $\omega_r=2\pi\times3.162$ GHz. We also measure the phases corresponding to the amplitudes of transmission spectra through the SAW resonator, the crossover from the anticrossing to the level attraction can also be observed.

In fact, our experiments show a temperature induced transition from the quantum to the classical regime in the cQAD system. In the lower temperature limit, e.g., the base temperature of the dilution refrigerator,  the thermal phonon number is near zero. As shown in Fig.~\ref{Figure5}(a), we observe a clear vacuum Rabi splitting, which corresponds to the distance between two peaks in curve A calculated using the transmon model of the lowest five energy levels or curve A$^{\prime}$ calculated using the transmon model of the lowest two energy levels of Fig.~\ref{Figure5}(g). With the increase of the temperature, the thermal excitations in the system are increased and more peaks should appear in the spectrum.  Although we can not observe multiple peaks in current experiment due to the limitation of sample parameters and measurement techniques, but as shown in the curve C$'$ in  either Fig.~\ref{Figure5}(g) or Fig.~\ref{Figure5} (h), two more peaks for the transitions from the state $|2,+\rangle$ to the state $|1,+\rangle$  and from  the state $|2,-\rangle$ to the state  $|1,-\rangle$ could appear. This shows a quantum nature of the system. If the temperature is further increased, and the thermal excitation number is far larger than one, all of the lower energy levels are occupied. The transitions occur between higher energy levels with the large phonon number $n$. In this case, the frequency differences between $\omega_{1,n}$ in Eq.~(\ref{Eq:7}) and $\omega_{2,n}$ in Eq.~(\ref{Eq:8}), which are proportional to $2g(\sqrt{n+1}-\sqrt{n})$,  becomes smaller than the decay rates of the acoustic resonator and qubit, i.e., $2g(\sqrt{n+1}-\sqrt{n}) << \Gamma,\,\kappa$, thus the peaks corresponding to these frequencies are not resolved and merge into one around the resonant frequency $\omega$, the height of this central peak is proportional to $\sqrt{n}$. Because the heights of the peaks corresponding to the frequencies $\omega_{3,n}$ and $\omega_{4,n}$ in the spectrum are proportional to $2g(\sqrt{n+1}-\sqrt{n})$, thus they are not observable in the limit of the large $n$. That is, all quantum nature disappear in the high temperature, the classical behavior is observed.  The numerical simulations in Fig.~\ref{Figure6}(i) clearly show the spectrum variations when the coupled system is changed from the quantum to classical regime by varying the environmental temperature. The change from the level splitting to the level attraction around $T_{c}=149$ mK indicates that the cross-over temperature of quantum-to-classical transition in our cQAD is about $T_{c}=149$ mK. When the temperature $T\gg149\,$mK, we observe the classical behavior of the system.

\subsection{Theoretical method and numerical simulations}
To further understand the quantum-to-classical transition induced by the temperature, we theoretically analyze and simulate the transmission spectra in Figs.~\ref{Figure6} (a), (c), (e) and (g) via the master equation
\begin{equation}
\begin{split}
\frac{\partial\rho}{\partial t}=&-i[\mathcal{H}+H_p,\rho]+\frac{\kappa}{2}\left[n_{\rm th}(\omega_r)+1\right]D[\hat{a}]\rho \\
&+\frac{\kappa}{2}n_{\rm th}(\omega_r)D[\hat{a}^\dagger]\rho \\
&+\frac{\Gamma}{2}\left[n_{\rm th}^{\prime}(\omega_a)+1\right]D\bigg[\sum_{l=e,f,\cdots}\frac{g_{l-1,l}}{g_{g,e}}\sigma_{l-1,l}\bigg]\rho \\
&+\frac{\Gamma}{2}n_{\rm th}^{\prime}(\omega_a)D\bigg[\sum_{l=e,f,\cdots}\frac{g_{l-1,l}}{g_{g,e}}\sigma_{l-1,l}^\dagger\bigg]\rho,
\end{split} \label{eq12}
\end{equation}
here $\mathcal{H}$ is ladder-model Hamiltonian with $l$ transmon levels~\cite{Koch2007}. In our theoretical study,  we use the Hamiltonian $\mathcal{H}$ with the lowest five energy levels of transmon for numerical simulation and the lowest two energy levels for theoretical analysis and comparison with the five energy levels, $\sigma_{l-1,l}=|l-1\rangle\langle l|$ and $\rho$  is the density matrix operator of the qubit-resonator system, $H_p=\varepsilon(\hat{a}^{\dagger}+\hat{a})/2$ is the interaction Hamiltonian between the probe field and the SAW resonator with the interaction strength $\varepsilon$, and $D[\hat{o}]\rho=2\hat{o}\rho{\hat{o}^\dagger}-{\hat{o}^\dagger}\hat{o}\rho-\rho{\hat{o}^\dagger}\hat{o}$ denotes the Lindblad type of the dissipation. $\kappa$ and $\Gamma$ are damping rates corresponding to the resonator and the transmon qubit, respectively. Noting that the temperature dependence of $\Gamma$ on piezoelectric substrate is ignored because it makes small difference to the results up to $349\,$mK in our experiment. In fact, the loss channels of qubits on piezoelectricity substrate are expected to be more complicate than qubits on Sapphire or Si~\cite{Lisenfeld2007,Sun2012,Diniz2020,Singh2020}. The related discussions are beyond the topic of this paper. The mean thermal phonon numbers for the resonator and the qubit are $n_{\rm th}(\omega_r)$ and $n_{\rm th}^{\prime}(\omega_a)$, respectively, which are calculated using formula of the Bose-Einstein distribution. The transmission rate, calculated analytically and numerically, is defined as
\begin{equation}
	t=\frac{i\kappa\langle{\hat{a}}\rangle_{ss}}{\varepsilon}.
\end{equation}

Before going to the numerical simulations, we first give a simple theoretical analysis. Using Hamiltonian $\mathcal{H}$ with the lowest two levels of transmon is enough for analysing characteristics of spectra. Under the secular approximation~\cite{Tian1992}, and with the detailed calculation steps given in the supplemental material~\cite{supplemental_material}, the rescaled transmission rate is given as 
\begin{eqnarray}
t\cdot\bigg(\frac{i \kappa}{\varepsilon}\bigg)^{-1}&=&\langle \hat{a} \rangle_{ss}=\frac{1}{\sqrt{2}}\sum_{\eta=\pm}\rho_{1,\eta;0}^{(1)}\\
&+&\sum_{\eta=\pm,\xi=\pm}\sum_{n=1}^{\infty} \frac{C^{n}_{\eta,\xi}}{2} \rho_{n+1,\eta;n,\xi}^{(1)}, \nonumber
\end{eqnarray}
with $C_{\eta,\xi}^n=\sqrt{n+1}+\epsilon_{\eta}\epsilon_{\xi}\sqrt{n}$ ($\epsilon_{\eta}=\pm1,\epsilon_{\xi}=\pm1$). The detailed expressions of $\rho_{1,\eta;0}^{(1)}$ and $\rho_{n+1,\eta;n,\xi}^{(1)}$, which are temperature-dependent, are given in supplemental material~\cite{supplemental_material}. The term of $\rho_{1,\eta;0}^{(1)}$ corresponds to the vacuum Rabi splitting. For given $n$, the relative strength of $C_{\pm,\pm}^n$ shows that amplitudes corresponding to transition frequencies $\omega_{1,n}$ and $\omega_{2,n}$ are larger than those corresponding to frequencies $\omega_{3,n}$ and $\omega_{4,n}$, and this results in that the peaks corresponding to $\omega_{3,n}$ and $\omega_{4,n}$ are difficult to be observed in the spectrum. In particular, the peaks corresponding to $\omega_{3,n}$ and $\omega_{4,n}$ are not observable in the limit of large $n$.  This result coincides with the conclusion discussed in Section \uppercase\expandafter{\romannumeral3}. For given $\eta$ and $\xi$, the relative strengths of $\rho_{1,\eta;0}^{(1)}$ and $C^{n}_{\eta,\xi} \rho_{n+1,\eta;n,\xi}^{(1)}$ corresponding to the heights of peaks in the spectrum varies when the environmental temperature $T$ is changed. In the lower temperature with the very smaller $n_{\rm th}$ and $n_{\rm th}^{\prime}$, the most obvious peaks correspond to those of vacuum Rabi splitting. While for the higher temperature with the larger $n_{\rm th}$ and $n_{\rm th}^{\prime}$, the most dominated peaks correspond to those of both vacuum Rabi splitting and the transitions from higher energy levels. When $n_{\rm th}$ and $n_{\rm th}^{\prime}$ become very large, all peaks merge into one peak, as shown in experimental result. We note that the result derived from the secular approximation does not agree well with the experiment, this is because all oscillatory terms are neglected as shown in supplemental material~\cite{supplemental_material}. These neglected terms have small effect on the result under the condition of the lower temperature. However, they take great effect on the result under the condition of the high temperature. This is because the effective dissipation becomes comparable or larger than the coupling strength between the SAW resonator and the qubit with the increase of the environmental temperature. Nevertheless, it is helpful enough to clearly show the variation trend of peaks in the spectrum with the increase of the temperature. More accurate result is shown below by numerical simulation.

Simulation on the transmission spectra is completed by numerically solving master equation in Eq.~(\ref{eq12}) under different environmental temperature~\cite{Rau2004}. The simulation results are shown in curves in Fig.~\ref{Figure5}(g), Figs.~\ref{Figure6}(b), (d), (f), (h) and Fig.~\ref{Figure6}(i), respectively. Figure~\ref{Figure5}(g) and Fig.~\ref{Figure6}(i) are
resonant transmission spectra at the different temperature.  We find that the experimental results at different temperatures can be well explained by numerically simulations. At the base temperature around $16.5$ mK, the anticrossing in the transmission spectrum, which is so called vacuum Rabi splitting in Fig.~\ref{Figure5}(a), demonstrates that the qubit is strongly coupled to SAW resonator. Thus two peaks appear in the curve A in Fig.~\ref{Figure5}(g). With the increase of the temperature, high energy levels are excited, the anticrossing becomes not very clear in Fig.~\ref{Figure5}(c), two more peaks occurs as shown in curve C$'$ of Fig.~\ref{Figure5}(g) or Fig.~\ref{Figure5} (h). When the temperature is increased to $T_{c}\sim\hbar\omega_{r}\approx149$ mK, around this temperature, anticrossing for the vacuum Rabi splitting gradually disappears as shown in Fig.~\ref{Figure6}(a) and the dominant peaks become wider. Above this temperature, all peaks in the transmission spectra start to merge into each other with the increase of the temperature as shown in Fig.~\ref{Figure6}(i). When the temperature becomes higher and higher, e.g., at the temperature $T\approx349$ mK which means $T\gg T_{c}$, two hybridized modes coalesce as shown in Fig.~\ref{Figure6}(g), thus there is only one peak at the frequency $\omega_r=2\pi\times3.162$ GHz as shown in curve corresponding to $T=349$ mK of Fig.~\ref{Figure6}(i).

\section{CONCLUSIONS}\label{conclusion}
In this work, we experimentally realize a cQAD system, which consists of a transmon qubit strongly coupled to both a 2D SAW resonator and a $1$D microwave transmission line. With the architecture, we demonstrate AC-Stark shift in weak dispersive system with the average phonon number inside SAW resonator in a wide range, which is from $\langle n\rangle\sim0.1$ to $\langle n\rangle\sim1000$. We show the propagating microwave field in the microwave transmission line can be controlled by phonons inside the SAW resonator with the average phonon number down to $\langle n\rangle\sim21$. This is the first step towards control of microwave photons by phonons in a quantum hybrid system on-chip. We also systematically study the quantum-to-classical crossover of the cQAD system. With the increase of the effective temperature of environment, we observe the crossover from level anticrossing to level attraction in transmission spectra. We expect that we could observe clear evidence of a quantum nature of the coupled system by two more peaks in resonant spectra when the environmental temperature is lower than the crossover temperature with improved architecture, e.g., a flux qubit is strongly coupled to a SAW resonator and a microwave transmission line simultaneously in future. Our study lays a solid foundation for further experiments in the context of quantum physics and information using cQAD system.

\begin{acknowledgments}
This work is supported by NSFC under Grant No.~61833010, No.~12074117 and No.~12061131011, Hunan Province Science and Technology Innovation Platform and Talent Plan (Excellent Talent Award) under Grant No.~2017XK2021. LZ is supported by the NSFC under Grant Nos.~11975095 and 11935006. LMK is supported by the NSFC under Grant No.~11775075 and No.~11434011. YXL is supported by the Key R\&D Program of Guangdong province under Grant No.~2018B030326001 and NSFC under Grant No.~11874037.
\end{acknowledgments}

\begin{center}
	\section{Supplemental Material}
\end{center}

\setcounter{equation}{0}
\setcounter{figure}{0}
\renewcommand{\thefigure}{S\arabic{figure}}
\renewcommand{\theequation}{S\arabic{equation}}
\setcounter{secnumdepth}{3}

\makeatletter
\def\@hangfrom@section#1#2#3{\@hangfrom{#1#2#3}}
\makeatother


\subsection{Splectrum of light transmitted by the cavity}

In this section, we show how temperature influences the anticrossing analytically following the method given in reference~\cite{Tian92}.

The system we discussed above can be modeled as J-C model. Under rotating wave approximation and dipole approximation, the master equation of our model can read
\begin{equation}
	\begin{split}
		\dot{\rho}=&-i[H+H_p,\rho]+\frac{\kappa}{2}[n_{th}(\omega_r)+1]D[\hat{a}]\rho+\frac{\kappa}{2}n_{th}(\omega_r)D[\hat{a}^\dagger]\rho+ \\
		&\frac{\Gamma}{2}[n_{th}^{\prime}(\omega_a)+1]D[\sigma_{-}]\rho+\frac{\Gamma}{2}n_{th}^{\prime}(\omega_a)D[\sigma_{+}]\rho
	\end{split}  \label{master_equation}
\end{equation}
where $\rho$  is the density matrix operator of the qubit-resonator system, $H_p=\varepsilon(\hat{a}^{\dagger}+\hat{a})/2$ is the interaction Hamiltonian between the probe field and the SAW resonator with the interaction strength $\varepsilon$, and $D[\hat{o}]\rho=2\hat{o}\rho{\hat{o}^\dagger}-{\hat{o}^\dagger}\hat{o}\rho-\rho{\hat{o}^\dagger}\hat{o}$ denotes the Lindblad type of the dissipation. Damping rates are related to the resonator decay rate $\kappa$, the intrinsic relaxation of the transmon qubit from excited state $|e\rangle$ to ground state $|g\rangle$ at rate $\Gamma$ and the creation of phonon as well as the excitement of the qubit energy level due to the thermal bath with the mean thermal phonon $n_{th}(\omega_r)$ and $n_{th}^{\prime}(\omega_a)$ which distribute under Bose-Einstein distribution $n_{th}=[\exp(-\hbar \omega_r / k_B T)-1]^{-1}$ and $n_{th}^{\prime}=[\exp(-\hbar \omega_a /k_B T)-1]^{-1}$.

It is difficult to obtain exact analytical solution of equation.(\ref{master_equation}) because
the difference equations contain infinite components of different frequencies, Thus, perturbation method is an useful method to obtain a solution which can elaborate our conclusion qualitatively. Regarding $H_p$ as perturbation, density matrix can be written as
\begin{equation}
	\rho=\rho^{(0)}+\rho^{(1)}+\rho^{(2)}+\rho^{(3)}+\cdots+\rho^{(n)}
\end{equation}
Here $\rho^{(0)}$ is the steady state solution of the coupled system without probing field. $\rho^{(n)}$ represents the $n$th order approximation of density matrix under perturbation. In our method, $\rho=\rho^{(0)}+\rho^{(1)}$ is enough to derive the analytical solution.

In the first-order approximation, we have
\begin{equation}
	\begin{split}
		&\dot{\rho}^{(0)}=-i[H, \rho^{(0)}]+\mathcal{L}[\rho^{(0)}] \\
		&\dot{\rho}^{(1)}=-i[H_p, \rho^{(0)}]-i[H, \rho^{(1)}]+\mathcal{L}[\rho^{(1)}]
	\end{split}
\end{equation}

First, $\rho_{0}$ is solved under secular approximation. As shown in main text, the eigenvectors of J-C Hamiltonian are dressed-states $|n,\pm \rangle$ and $|g,0 \rangle$ whose corresponding frequencies are $\omega_{n,\pm}$ and $\omega_g$. Under strong coupling condition, we expand master equation (\ref{master_equation}) under dressed-states basis using the secular approximation. We make the transformation
\begin{equation}
	\rho=e^{-iHt} \bar{\rho} e^{iHt}  \label{transmation}
\end{equation}

We can drop all oscillatory terms in the transformed master equation using secular approximation. Noting that this approximation requires that all peaks in spectrum are well separated compared with their linewidth, it may not be a good approximation when the transition are mostly from higher energy level.  Nevertheless, it is useful to obtain the analytical results which is useful to analyse the influence of temperature on anticrossing qualitatively. In secular approximation, the diagonal matrix elements of $\bar{\rho}^{(0)}$ read

\begin{equation}
	\begin{split}
			&\dot{\bar{\rho}}_0^{(0)}=-2\gamma_{g,1} \bar{\rho}_0^{(0)} + 2\gamma_{1,g}\bar{\rho}_1^{(0)} \\
	    	&\dot{\bar{\rho}}_1^{(0)}=-(\gamma_{1,g}+2\gamma_{1,2})\bar{\rho}_1^{(0)} + \gamma_{g,1} \bar{\rho}_0^{(0)} + 2\gamma_{2,1} \bar{\rho}_2^{(0)} \\
		    &\dot{\bar{\rho}}_n^{(0)}=-2(\gamma_{n,n-1} + \gamma_{n,n+1}) \bar{\rho}_n^{(0)} + 2\gamma_{n-1,n}\bar{\rho}_{n-1}^{(0)} + 2\gamma_{n+1,n}\bar{\rho}_{n+1}^{(0)}
	\end{split} \label{Eqs4}
\end{equation}
where matrix element $\bar{\rho}_0^{(0)}=\langle g,0 | \bar{\rho}^{(0)} |g,0 \rangle$, $\bar{\rho}_n^{(0)}=\langle n,+|\bar{\rho}^{(0)}|n,+\rangle=\langle n,-|\bar{\rho}^{(0)}|n,-\rangle$, the transition rates are
\begin{equation}
	\begin{split}
			&\gamma_{g,1}=(\Gamma /2) n_{th}^{\prime} + (\kappa /2) n_{th}  \\
		    &\gamma_{1,g}=(\Gamma /2) (n_{th}^{\prime}+1) + (\kappa /2) (n_{th}+1)  \\
		    &\gamma_{n,n+1}=(\Gamma /4) n_{th}^{\prime} + [(\kappa /4) n_{th}] (2n+1)   \qquad n=1,2,\cdots\\
		    &\gamma_{n,n-1}=(\Gamma /4) (n_{th}^{\prime}+1) + [(\kappa /4) (n_{th}+1)](2n-1)  \qquad n=2,3,\cdots
	\end{split}
\end{equation}
For off-diagonal matrix elements, the equations of motion without secular approximation read
\begin{equation}
	\begin{split}
		\dot{\bar{\rho}}_{1,+;0}^{(0)}=&-\beta \bar{\rho}_{1,+;0}^{(0)} + [\frac{\Gamma}{4} - (2n_{th}+1)\frac{\kappa}{4}] \bar{\rho}_{1,-;0}^{(0)} e^{-i(\omega_{1,-}-\omega_{1,+})t} + \\
		&\frac{1}{\sqrt{2}} \bigg\lbrace \bigg[\frac{\Gamma}{2}(n_{th}^\prime+1)+(\sqrt{2}+1)(n_{th}+1)\frac{\kappa}{2}\bigg]\bar{\rho}_{2,+;1,+}^{(0)} e^{-i(\omega_{2,+}+\omega_{g}-\omega_{1,+}-\omega_{1,+})t} \\
		&+  \bigg[\frac{\Gamma}{2}(n_{th}^\prime+1)+(\sqrt{2}-1)(n_{th}+1)\frac{\kappa}{2}\bigg]\bar{\rho}_{2,-;1,-}^{(0)} e^{-i(\omega_{2,-}+\omega_{g}-\omega_{1,+}-\omega_{1,-})t} \\
		&+ \bigg [-\frac{\Gamma}{2}(n_{th}^\prime+1)+(\sqrt{2}+1)(n_{th}+1)\frac{\kappa}{2}\bigg]\bar{\rho}_{2,+;1,-}^{(0)} e^{-i(\omega_{2,+}+\omega_{g}-\omega_{1,+}-\omega_{1,-})t}\\
		& + \bigg[-\frac{\Gamma}{2}(n_{th}^\prime+1)+(\sqrt{2}-1)(n_{th}+1)\frac{\kappa}{2}\bigg]\bar{\rho}_{2,-;1,+}^{(0)} e^{-i(\omega_{2,-}+\omega_{g}-\omega_{1,+}-\omega_{1,+})t} \bigg\rbrace
	\end{split} \label{eqwithout}
\end{equation}
The oscillatory frequency $\omega_{n,\pm}=(n-1/2)\omega\pm\sqrt{n}g$ as well as $\omega_g=-\omega/2$ when the transmon qubit resonantly interacts with the SAW resonator, we can drop all oscillatory terms using secular approximation in order to get the analytical expression, the equation above can be simplied as
\begin{equation}
	\begin{split}
		\dot{\bar{\rho}}_{1,\eta;0}^{(0)}&=-\beta \bar{\rho}_{1,\eta;0}^{(0)} \\
		\dot{\bar{\rho}}_{2,\eta;1,\xi}^{(0)}&=-\alpha_1 \bar{\rho}_{2,\eta;1,\xi}^{(0)} \\
		\dot{\bar{\rho}}_{n+1,\eta;n,\xi}^{(0)}&=-\alpha_n \bar{\rho}_{n+1,\eta;n,\xi}^{(0)}
	\end{split} \label{eqs7}
\end{equation}
where $\bar{\rho}_{1,\eta;0}^{(0)}=\langle 1,\eta|\bar{\rho}^{(0)}|g,0\rangle$, $\bar{\rho}_{n+1,\eta;n,\xi}^{(0)}=\langle n+1,\eta|\bar{\rho}^{(0)}|n,\xi \rangle$, $\eta$ and $\xi$ are either + or -. The coefficient $\beta$ and $\alpha_n$ read
\begin{equation}
	\begin{split}
		&\beta=\gamma_{1,g}/2 + \gamma_{1,2} + \gamma_{g,1} \\
		&\alpha_1= \gamma_{2,1} + \gamma_{2,3} + \gamma_{1,2} +  \gamma_{1,g}/2 \\
		&\alpha_n=\gamma_{n+1,n}+\gamma_{n+1,n+2}+\gamma_{n,n+1}+\gamma_{n,n-1} \quad n=2,3\cdots
	\end{split}
\end{equation}
Under transformation.(\ref{transmation}), the relation of original matrix elements and transformed matrix elements reads
\begin{equation}
	\begin{split}
		&\rho_{0}^{(r)}=\bar{\rho}_0 ^{(r)} \qquad  \quad  \rho_{n}^{(r)}=\bar{\rho}_n^{(r)}  \\
		&\rho_{1,\eta;0}^{(r)}=exp[-i(\omega+\epsilon_{\eta}g)t] \bar{\rho}_{1,\eta;0}^{(r)} \\
		&\rho_{n+1,\eta;n,\xi}^{(r)}=exp[-i(\omega+\epsilon_{\eta}\sqrt{n+1}g-\epsilon_{\xi}\sqrt{n}g)t]\bar{\rho}_{n+1,\eta;n,\xi}^{(r)}
	\end{split} \label{eqs8}
\end{equation}
where $r$ refers to $r$th order approximation and $\epsilon_{\eta}(\epsilon_{\xi})=1,-1$ for $\eta(\xi)=+,-$.

Setting the left side of equation.(\ref{Eqs4}) to zero, the steady-state population of the J-C Hamiltonian can be obtained
\begin{equation}
	\begin{split}
		&\bar{\rho}_1^{(0)ss}=(\gamma_{g,1}/\gamma_{1,g}) \bar{\rho}_g^{(0)ss} \\
		&\bar{\rho}_{n+1}^{(0)ss}=(\gamma_{n,n+1}/\gamma_{n+1,n}) \bar{\rho}_n^{(0)ss}
	\end{split}
\end{equation}

Then the steady-states populations write
\begin{equation}
	\rho_{n}^{(0)ss}=\bar{\rho}_n^{(0)ss}=\prod_{m=1}^{n}\bigg(\frac{\Gamma n_{th}^{\prime} + \kappa n_{th} (2m-1)}{\Gamma (n_{th}^{\prime}+1) + \kappa (n_{th}+1)(2m-1)}\bigg)\rho_g^{(0)ss}    \label{eqs11}
\end{equation}
With the condition that
\begin{equation}
	2\sum_{n=1}^{\infty}\rho_{n}^{(0)ss}+\rho_{g}^{(0)ss}=1 \label{eqs12}
\end{equation}

Due to $\rho_{1,\eta;0}^{(0)ss}=0, \rho_{n,\eta;n+1,\xi}^{(0)ss}=0$ by solving equation.(\ref{eqs7}).  $\rho^{(1)}$ is needed to be solved in order to attain non-zeros non-diagnol matrix elements. According to equation.(\ref{eqs8}), $\dot{\rho}_{1,\eta;0}^{(1)}=-i\omega_{1,\eta;0} \cdot \exp[-i\omega_{1,\eta;0}t] \bar{\rho}_{1,\eta;0}^{(1)}+\exp[-i\omega_{1,\eta;0}t] \dot{\bar{\rho}}_{1,\eta;0}^{(1)}$ and $\dot{\rho}_{n+1,\eta;n,\xi}^{(1)}=-i\omega_{n+1,\eta;n,\xi} \cdot \exp[-i\omega_{n+1,\eta;n,\xi}t] \bar{\rho}_{n+1,\eta;n,\xi}^{(1)}+\exp[-i\omega_{n+1,\eta;n,\xi}t] \dot{\bar{\rho}}_{n+1,\eta;n,\xi}^{(1)}$, where $\omega_{1,\eta;0}=\omega_0+\epsilon_{\eta}g,  \omega_{n+1,\eta;n,\xi}=\omega_0+\epsilon_{\eta}\sqrt{n+1}g-\epsilon_{\xi}\sqrt{n}g^{(1)}$ for $n\ge1$, this leads to the steady states of transformed non-diagnol matrix elements $\dot{\bar{\rho}}_{1,\eta;0}^{(1)}=i\omega_{1,\eta;0} \cdot \bar{\rho}_{1,\eta;0}^{(1)}$ and $\dot{\bar{\rho}}_{n+1,\eta;n,\xi}^{(1)}=i\omega_{n+1,\eta;n,\xi} \cdot \bar{\rho}_{n+1,\eta;n,\xi}^{(1)}$

Using the same method, the steady states equation of density matrix $\rho_{1,\eta;0}^{(1)}$ and $\rho_{n+1,\eta;n,\xi}^{(1)}$ reads

\begin{equation}
	\begin{split}
		&[i(\omega_{1,\eta;0})+\beta]\rho_{1,\eta;0} ^{(1)}=-i\frac{\varepsilon}{\sqrt{2}}[\rho_{0}^{(0)}-\rho_{1}^{(0)}]   \\
		[i(\omega_{n+1,\eta;n,\xi})&+\alpha_n]\rho_{n+1,\eta;n,\xi}^{(1)}=-i\frac{(\sqrt{n+1}+\epsilon_{\eta}\epsilon_{\xi}\sqrt{n})\varepsilon}{2}[\rho_{n}^{(0)}-\rho_{n+1}^{(0)}]
	\end{split} \label{eqs16}
\end{equation}
The steady states value of $\langle \hat{a} \rangle$ reads
\begin{equation}
	\begin{split}
		\langle \hat{a} \rangle_{ss}&=\frac{1}{\sqrt{2}} \sum_{\eta=\pm} \langle \hat{S}_{0\eta}\rangle+\sum_{\eta=\pm,\xi=\pm} \sum_{n=1}^{\infty} \frac{C^{n}_{\eta,\xi}}{2}\langle \hat{S}_{\xi ,\eta}^n \rangle \\
		&=\frac{1}{\sqrt{2}} \sum_{\eta=\pm} \rho_{1,\eta;0}^{(1)}+\sum_{\eta=\pm,\xi=\pm} \sum_{n=1}^{\infty} \frac{C^{n}_{\eta,\xi}}{2} \rho_{n+1,\eta;n,\xi}^{(1)}
	\end{split} \label{eqs17}
\end{equation}
The operator $\hat{S}_{0\eta}=|g,0\rangle \langle 1,\eta|$, $\hat{S}_{\xi  \eta}^n=|n,\xi \rangle \langle n+1,\eta|$ with $C_{\eta,\xi}^n=\sqrt{n+1}+\epsilon_{\eta}\epsilon_{\xi}\sqrt{n}$. Transmission spectrum $t$ in main text can be obtained by combining equation equations.(\ref{eqs11}), (\ref{eqs12}), (\ref{eqs16}) and (\ref{eqs17}).

As shown above, coefficient $C_{\pm,\pm}^{n}$ describe the relative strength of transition frequency for the same
$n$ in spectrum, the amplitude corresponding to frequency $\omega_{1,n}$ and $\omega_{2,n}$ is much larger than frequency $\omega_{3,n}$ and $\omega_{4,n}$. And the relative strength $C^{n}_{\eta,\xi}\rho_{n+1,\eta;n,\xi}^{(1)}$ which is temperature-dependent for same $\eta,\xi$ indicates the resonant peak which is dominant in spectrum.For increasing temperature $T$ (corresponding to larger $n_{th}$ and $n_{th}^{\prime}$), the most distinct two peaks in spectrum become more and more close which corresponding to frequency $\omega_{1,n}$ and $\omega_{2,n}$ with larger $n$. Finally, the two peaks can not be  distinguished and merged into one peak when temperature is large enough. For small $T$, secular approximation can be applied without any problem. However, when $T$ increase, the effective decay rate $\Gamma_{eff}$, $\kappa_{eff}$ become $\Gamma(n_{th}^\prime+1)$ and $\kappa(n_{th}+1)$, the restriction of secular approximation $g\gg \Gamma_{eff},\kappa_{eff}$ don' t manifest in this system due to the neglected oscillatory terms in equation.(\ref{eqwithout}) makes impact. This results in the deviation between analytical and numerical results. However, analytical result still shows the change of dominant peaks qualitatively. For larger $T$,  the dominant energy level $n$ corresponding to largest  $C^{n}_{\eta,\xi}\rho_{n+1,\eta;n,\xi}^{(1)}$ get larger, which means the dominant peaks become close to central frequency and finally merge into one peak, this tendency agrees well with numerical result. For further exact result, see numerical result which including oscillatory terms in main text.

\end{document}